\documentclass[preprint,showpacs,preprintnumbers,amsmath,amssymb]{revtex4}
\usepackage{graphicx}% Include figure files
\usepackage{dcolumn}% Align table columns on decimal point
\usepackage{bm}% bold math

\begin{document}
\title{Bending-Filament Model for the Buckling and Coiling Instability \\of Viscous Fluid Rope}
\author{Shin-ichiro Nagahiro}
 \affiliation{%
Department of Mechanical Engineering, Miyagi National College of Technology, Miyagi 981-1239, Japan
}%
\author{Yoshinori Hayakawa}
\affiliation{%
Department of Physics, Tohoku University, Aoba-ku, Sendai, 980-8578, Japan
}%
\date{\today}% It is always \today, today,
\begin{abstract} 
A simple model is proposed for the buckling and coiling instability of a viscous ``fluid rope" falling on a plane. By regarding a fluid rope as a one-dimensional flow, this model accounts for only the axial and shared viscous forces. Our model successfully reproduces several experiments with no adjustable parameters, such as the existence of three distinct coiling regimes reported in Phys. Rev. Lett. {\bf 93}, 214502 (2004). Our model allows for the discussion of unsteady motion. An expression for the critical fall height at which the coiling frequency changes from a decrease to increase was phenomenologically derived. It was found that the coil-uncoil transition shows remarkable hysteresis only for weak gravity condition.  
 
\end{abstract}
\maketitle
In the low Reynolds number regime, the dynamics of viscous fluid confined 
in rigid boundaries do not exhibit complex behavior because
the possible eigenmodes are limited.
However, if the fluid interface can move freely  
and deform largely, this is not the case, and rich variety of dynamics can emerge
{\it below} a critical Reynolds number.
Backing and coiling instabilities of fluid jets are examples of such systems,
and have been studied for several decades in the laboratory
\cite{barnes1, barnes2,cruick2,mahadevan, ribe_prl, ribe_pre}.
A fluid rope and coiling can also be observed in daily life, such as when honey is poured from a teaspoon onto toast.

Although the mathematical treatment of a largely deformed fluid is not theoretically straightforward,
the conditions for the onset of coiling instability have been thoroughly examined 
in terms of linear stability analysis with some simplifications and assumptions \cite{cruick3, yarin, ribe_PF}.

Recently, Ribe derived the differential equations to describe steady state coiling for a very thin rotating rope, and showed the numerical solutions for various conditions \cite{ribe_prsl}. His analysis reveals that the steady solution is multi-valued so that there might be a discontinuity in the selected coiling frequency.

Ribe also demonstrated the existence of three distinct coiling modes;
viscous, gravitational, and inertial regimes.
Phenomenologically, one could understand that buckling and coiling instability 
occurs under the mechanical balance
between the driving force of a steady flow and the internal stress due to viscosity.
Signifying the magnitudes of force for fluid injection as $F_P$, 
the gravitational force acting to fluid as $F_G$, the inertial force as $F_I$,
and the yielding stress due to viscosity as $F_V$,
the viscous coiling regime corresponds the condition $F_P \sim F_V$,
the gravitational coiling regime $F_G \sim F_V$, and 
the inertial coiling regime $F_I \sim F_V$, respectively.
Dimensional analysis allows for the estimation of the coiling frequencies for each regime as
\begin{eqnarray}
\Omega_V & = & \frac{Q}{H a^2} \label{eq:viscous},\\
\Omega_G & = & \left( \frac{g Q^3}{\nu a^8} \right)^{1/4} \label{eq:gravitational},\\
\Omega_I & = & \left( \frac{Q^4}{\nu a^{10}} \right)^{1/3} \label{eq:inertial},
\end{eqnarray}
where $Q$ is the flow rate, $H$ is the fall height, $\nu$ is the kinetic viscosity,
$g$ is the gravitational acceleration, and $a$ is the radius of the rope.

Although these studies seem to succeed in giving the onset of coiling
with infinitesimal amplitude and steady coiling frequencies, 
in order to discuss the stability and further turbulent states,
a numerical model is still needed that can describe the entire dynamics including
transient states with fewer computational costs. 

In this Letter, an alternative numerical model for a small deflection in a fluid rope is proposed and coiling frequency as a function of fall height is investigated. In addition, the coil-uncoil transition and its hysteresis effect are discussed, as well as the transition height from viscous to gravitational coiling.

%%%%%%%%%%%%%%% Figure1 %%%%%%%%%%%%%%%%%
\begin{figure}[tb]
\begin{center}
\includegraphics[width= 3.6cm, keepaspectratio]{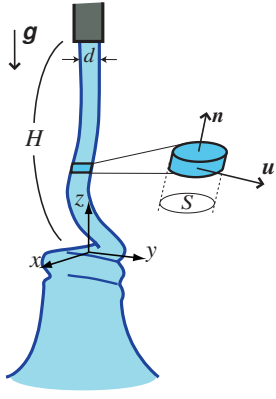}
\caption{A schematic view of a fluid rope coiling.}
 \label{fig:model}
\end{center}
\end{figure}

Figure \ref{fig:model} is a schematic view of a fluid rope injected from an orifice at a sufficient height. We fix the origin of the reference flame at the point where the rope begins to coil and align the $z$ axis parallel to gravity $\bm g$. In the present model, we require a ``steady coiling state" in which the upward growth speed of the coil is balanced by the speed of downward flow at the top of coil, i.e. the point at which the rope changes to the coil does not move.

We consider a uniform flow within the rope and write the flow velocity as $\bm w=w\bm n$, where $\bm n=(n_x, n_y, n_z)$ is the unit vector tangential to the rope. Since we restrict the present model to the case of slight deflection,  
the vector $\bm n$ is almost parallel to the $z$ axis. We take account of only the first order of $n_x$ and $n_y$, and replace the derivative with respect to $\bm n$ by the derivative with respect to $z$. In this manner, the axial stress is written as $\sigma_n=3\eta \partial w/\partial z$, where $3\eta$ is the extension (or compression) viscosity \cite{trouton}. Hereafter, we scale flow speed with the injection speed at the nozzle $w_{\rm in}$ and length with the orifice diameter $d$. Then, the dimensionless equation of motion for $w$ is
\begin{equation}
\left(\frac{\partial}{\partial t}+w\frac{\partial}{\partial z}\right)w=\frac{3}{S {\rm Re}}\frac{\partial}{\partial z}\left(S\frac{\partial w}{\partial z}\right) -\frac{1}{\rm Fr},
 \label{eq:z_flow}
\end{equation}
where ${\rm Re}=d|w_{\rm in}|/\nu$ is the Reynolds number and ${\rm Fr}=w_{\rm in}^2/gd$ is the Froude number. $S$ is the rope's cross-sectional area taken to be parallel to $xy$ plane. We determine $S$ by solving the following continuity equation
\begin{equation}
\frac{\partial S}{\partial t}= \frac{\partial}{\partial z}(Sw).
 \label{eq:continuity}
\end{equation}
In order to describe the deformation of the fluid rope, we divide the rope into thin cylindrical  elements whose motion is restricted in the $xy$ plane, as shown in Fig. \ref{fig:model}. Denoting the position of the element $\bm q=(q_x, q_y)$ and velocity $\bm u=(u_x,u_y)$, these quantities obey
\begin{equation}
\left(\frac{\partial}{\partial t}+w\frac{\partial}{\partial z}\right)q_i=u_i,
\label{eq:eqofmo_1}
\end{equation}
where $i=x$ and $y$. The total stress acts on the cylindrical element is the sum of viscous shear stress $\sigma_i = \eta\partial(u_i+wn_i)/\partial z$ and the $i$th component of the axial stress $n_i\sigma_n$. Therefore, we obtain the equation of motion for a cylindrical element as 
\begin{equation}
\left(\frac{\partial}{\partial t}+w\frac{\partial}{\partial z}\right)u_i=
\frac{1}{S{\rm Re}}\frac{\partial}{\partial z}
\left\{S\left(\frac{\partial u_i}{\partial z}+4n_i\frac{\partial w}{\partial z}+w\frac{\partial n_i}{\partial z}\right)\right\}
\label{eq:eqofmo_2}
\end{equation}

Next, we discuss the boundary conditions of the present model. At the injection point $z=H$, the fluid rope is fixed, or
\begin{equation}
q_i(H)=0,~~u_i(H)=0,
 \label{eq:}
\end{equation}
and $S(H)=\pi/4$.
We neglect the relaxation of Poiseuille flow to plug flow in the neighborhood of the orifice, and assume a uniform flow at the injection point as $w(H)=-1$. Let us turn to the boundary conditions at $z=0$. We assume that
\begin{equation}
q_i'(0)=0,~~u_i'(0)=0, \label{eq:bc_2}
\end{equation}
where a prime indicates a derivative with respect to $z$. Let $w_{\rm max}$ be the maximum of $|w(z)|$. Experimentally, Cruickshank and Munson found a relation 
$w_b\simeq -0.13w_{\rm max}$, where $w_b$ is the downward flow speed at the top of coiling region \cite{cruick1}. This relation is fairly independent of flow conditions such as the viscosity of the liquid, the orifice diameter or the fall height. The velocity $w(0)$ in the present model corresponds to $w_b$, we therefore require the following time dependent boundary condition
\begin{equation}
w(0) = -0.13 w_{\rm max}.\label{eq:bc_bottom}
\end{equation}

We numerically solve the equations (\ref{eq:z_flow}), (\ref{eq:continuity}), (\ref{eq:eqofmo_1}) and (\ref{eq:eqofmo_2}) setting an axial flow,  
$w(z)=-1$, $S(z)=\pi/4$, $q_i(z)=0$, and $u_i(z)=0$ as an initial condition. For $q_i(z)$, a small roughness with amplitude of $0.01$ is given.
The tangent vector $\bm n$ is calculated from the derivative of $\bm q$ with respect to $z$ as $\bm n = \left({\partial q_x}/{\partial z} , {\partial q_y}/{\partial z}, 1\right)/C$ with $C=\sqrt{(\partial q_x/\partial z)^2+(\partial q_y/\partial z)^2+1}$. 
The control parameters of the numerical calculations are ${\rm Re}$, ${\rm Fr}$ and the fall height $H$. 
%%%%%%%%%%%%%%% Figure2 %%%%%%%%%%%%%%%%%
\begin{figure}[tb]
\begin{center}
\includegraphics[width= 7.5cm, keepaspectratio]{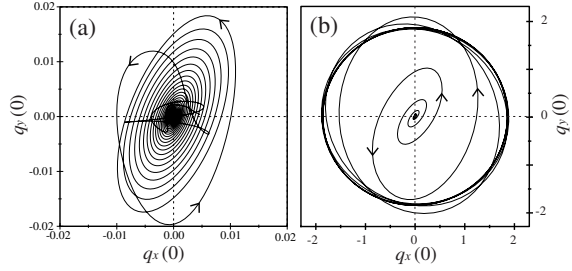}
\caption{Trajectories of the bottom of the model rope starts at $t=0$: (a) Trajectory in the case that axial flow is stable at the steady state (${\rm Re}=10.0$, ${\rm Fr}=100.0$, $H/d=10$). (b) Trajectory of circular coiling (${\rm Re}=1.0$, ${\rm Fr}=1.0$, $H/d=10$). 
 }
 \label{fig:orbits}
\end{center}
\end{figure}

Figure \ref{fig:orbits} shows trajectories of $\bm q$ in the plane $z=0$. Simulation movies are available\cite{mov}. In the case that $\rm Re$ is sufficiently large, the axial stagnation flow is stable, thus the trajectory converges to the origin, as shown in Fig. \ref{fig:orbits}(a). Such an axial flow could appear under no gravity condition. When we set $\rm Re$ smaller than a critical value, the flow starts to oscillate and a circular trajectory would appear at the steady state (Fig. \ref{fig:orbits}(b)).

%%%%%%%%%%%%%%% Figure3 %%%%%%%%%%%%%%%%%
\begin{figure}[tb]
\begin{center}
\includegraphics[width= 6.0cm, keepaspectratio]{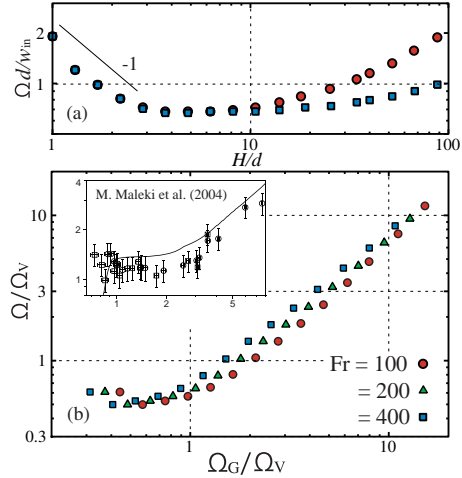}
\caption{Transition from viscous to gravitational coiling for ${\rm Fr}=100$, 200 and 400. (a) Dimensionless frequency $\Omega d/w_{\rm in}$ versus height in condition of ${\rm Re}=3$. (b) Frequency-height curve rescaled using the $\Omega_V$ and $\Omega_G$. The inset shows the experimental result by M. Maleki {\it et. al.} \cite{ribe_prl}.}
 \label{fig:viscous}
\end{center}
\end{figure}

Most of experimental works on fluid rope coiling discuss coiling frequency as a function of the fall height \cite{barnes1, barnes2, cruick2, ribe_prl, ribe_pre} or the rope's radius \cite{mahadevan}. Let us compare the coiling frequency of the present model with the experiments. To investigate the viscous coiling regime as a first step, Fr is set to a very large value, since the viscous coiling takes place when the effect of gravity is small. Figure \ref{fig:viscous}(a) shows a dimensionless frequency $\Omega d/w_{\rm in}$ as a function of fall height. We found that $\Omega \propto H^{-1}$ for lower fall heights, which agrees with the scaling relation in viscous coiling. For $H/d\simeq 10$, the frequency is almost constant. This behavior corresponds to the gravitational coiling, because the rope's radius hardly depends on the fall height. In figure \ref{fig:viscous}(b), the frequency-height curves are rescaled with $\Omega_V$ and $\Omega_G$ in order to compare the results with the experiment by M. Maleki {\it et. al.} (see the inset). The transition from viscous to gravitational coiling occurs at $\Omega_G/\Omega_V\simeq 1$. 
These results are in good agreement with the experiments including the transient regimes. For a higher fall height of $H\simeq10^2$, the frequency increases as $\Omega\propto H$, which agrees with the earliest observation by G. Barnes and J. MacKenze \cite{barnes1, barnes2}.

Supposing strong stretching $a\ll d$ due to gravity, the scaling relation Eq. (\ref{eq:inertial}) predicts $\Omega\propto H^{10/3}$ \cite{ribe_pre}. In this case, we reasonably choose a characteristic time scale $\sqrt{d/g}$. According to simple dimensional arguments \cite{cruick2}, frequency $\Omega\sqrt{d/g}$ must have the form:
\begin{equation}
\Omega\sqrt{d/g}=f\left(\frac{\nu Q}{gd^4}, \frac{gd^3}{\nu^{2}}, \frac{H}{d}\right),\label{eq:omega_scale}
\end{equation}
where $\nu=\eta/\rho$ is the kinematic viscosity. Here it is noticed that the first parameter in this function is the ratio of the viscous force to the gravitational force $F_V/F_G$, the second is the ratio of the gravitational force times the inertia force to the viscous force $F_GF_I/{F_V}^2$, and the third is the slenderness ratio. Because the strong stretching condition requires $F_V/F_G \lesssim 1$ and $H/d\gg1$, the inertial coiling regime appears in the condition:
\begin{equation}
\frac{\nu Q}{gd^4}\sim \frac{gd^3}{\nu^{2}} \lesssim1,~\frac{H}{d}\gg1
\end{equation}
Experimentally, Cruickshank and Munson first found $\Omega\propto H^2$ for $H/d$ higher than $10$, in the condition of $\nu Q/gd^4\simeq 0.1$-$1.7$, $gd^3/\nu^2\simeq0.1$ \cite{cruick2}. Ribe also observed $\Omega\propto H^{2.5}$ for $gd^3/\nu^2\simeq10^{-3}$ and $\nu Q/gd^4\simeq 0.5$ \cite{ribe_prl}.
 
\begin{figure}[tb]
\begin{center}
\includegraphics[width= 6.0cm, keepaspectratio]{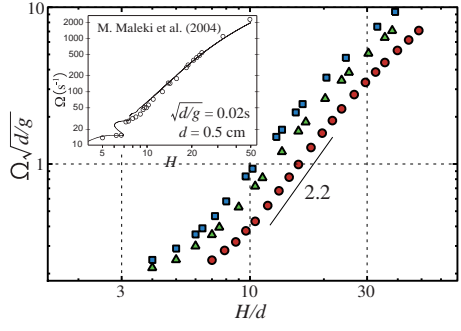}
\caption{Dimensionless frequency $\Omega \sqrt{d/g}$ versus height obtained from the present model. Parameters are: 
$\nu Q/gd^4=0.2$ and $gd^3/\nu^2=0.2$ (open circle), 
$\nu Q/gd^4=0.3$ and $gd^3/\nu^2=0.3$ (triangle) and 
$\nu Q/gd^4=0.2$ and $gd^3/\nu^2=0.5$ (rectangle). 
The inset shows the experimental result by M. Maleki {\it et. al.} \cite{ribe_prl}.}
 \label{fig:inertial}
\end{center}
\end{figure}

As shown in Fig. \ref{fig:inertial}, the present model gives $\Omega \propto H^{2.2}$ for the similar conditions. These experimental and theoretical results are successfully consistent. On the other hand, in the gravitational to inertial transitional range, the experiment shows a remarkable discontinuous jump in the frequency (inset in Fig. \ref{fig:inertial}). The present model, though, does not reproduce this behavior.

%%%%%%%%%%%%%%% Figure4 %%%%%%%%%%%%%%%%%

\begin{figure}[t]
\begin{center}
\includegraphics[width=5.5cm, keepaspectratio]{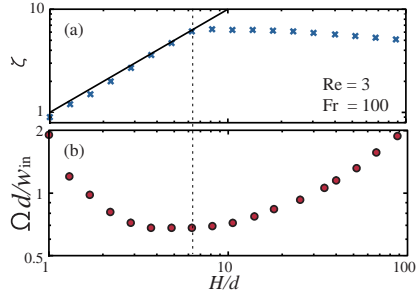}
\caption{(a) Numerically calculated value of $z_c$ for Re $=3$ and Fr $=100$. The solid line is the plot of $\zeta=H$. (b) Same data with Fig. \ref{fig:viscous} for comparison.}
 \label{fig:bp}
\end{center}
\end{figure}

Next, let us discuss critical fall height $H^*$, around which coiling frequency changes from decrease to increase (transition from viscous coiling to gravitational coiling). In figure \ref{fig:viscous}(a), the transition occurs at $H^*\simeq6d$. However, the $H^*$ should generally be a function of viscosity and gravitational acceleration. For the first step, we focus on the position $z=\zeta$ at which the flow velocity $w$ reaches a maximum. 
Note that, internal stress of fluid rope changes from tensile to compressive at this point. 
After a flow reaches steady state, it is observed that $\zeta$ hardly depends on time, even though the rope is oscillating. In Fig. \ref{fig:bp}(a), we plot $\zeta$ at steady state as a function of fall height. For a shorter fall height, the $\zeta$ is equal to $H$. This indicates that the maximum of $w$ appears at the point of injection and the fluid rope is wholly compressed. 
Comparing $\zeta$ with the frequency $\Omega d/w_{\rm in}$ shown in Fig.\ref{fig:bp}(b), we can find that the frequency decreases only when the relation $\zeta=H$ occurs. Therefore, viscous coiling appears in the case in which the whole the rope is compressed.
Because this feature can be seen for a wide range of Re and Fr as long as the viscous coiling regime exists, we claim that the $H^*$ is identical to the maximum of $\zeta$. The maximum of $\zeta$ can be realized as a relaxation length of which the effect of boundary condition at $z=0$ can travel through a rope. Using $g$ and $\nu$, we can uniquely construct a dimension of length as $g^{-1/3}\nu^{2/3}$. 
Therefore,
\begin{equation}
H^*=\max(\zeta)\propto g^{-1/3}\nu^{2/3}, \label{eq:scale}
\end{equation}
which is fairly close to the numerically calculated exponents $\max{(\zeta)}\propto g^{-0.36}\nu^{0.86}$.

%%%%%%%%%%%%%%% Figure5 %%%%%%%%%%%%%%%%%
%%%%%%%%%%%%%%% Figure6 %%%%%%%%%%%%%%%%%
\begin{figure}[t]
\begin{center}
\includegraphics[width= 6.0cm, keepaspectratio]{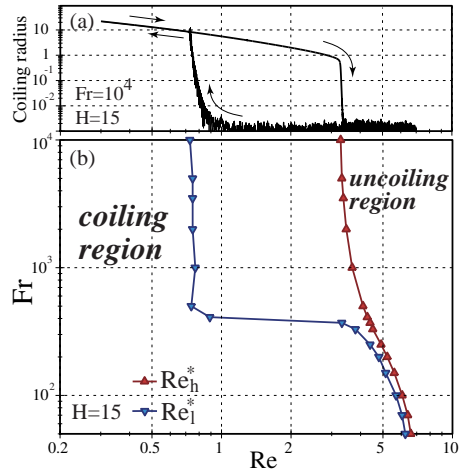}
\caption{(a) Numerically calculated coiling radius for ${\rm Fr=10^4}$ and $H=15$ with increasing and decreasing Re. (b) The coiling regime in Re-Fr plane for increasing ($\bigtriangleup$) and decreasing ($\bigtriangledown$) Reynolds number.}
 \label{fig:phase}
\end{center}
\end{figure}

Finally, we refer to a new and unforeseen result obtained from the present model. It is empirically known that fluid ropes do not coil if the Reynolds number of the flow is sufficiently large. We investigated the upper limit of Reynolds number of the present model and found that the transition occurs with remarkable hysteresis. In the simulation, we continuously add perturbation of amplitude $10^{-6}d$ for $q_i$. Fixing Fr and $H$, the value of Re is slowly increased or decreased not faster than 0.1\% change per one cycle of coiling. Figure \ref{fig:phase}(a) shows the coil-uncoil transition for weak gravity condition (Fr=$10^4$). For increasing Re, the coiling radius sharply falls down at ${\rm Re}\simeq 3.3 (\equiv{\rm Re}_l^*)$ and for decreasing Re, the radius somewhat gradually rises at ${\rm Re}\simeq 0.7(\equiv{\rm Re}_h^*)$.
Axial flow is absolutely stable at ${\rm Re}>{\rm Re}_h^*$, and is absolutely unstable at ${\rm Re}<{\rm Re}_l^*$. In the intermediate region (${\rm Re}_l^*<{\rm Re}<{\rm Re}_h^*$), fluid rope starts to coil if the amplitude of the perturbation is of order unity.
In Fig. \ref{fig:phase}(b), we show the coiling regime in Re-Fr plane for increasing and decreasing Reynolds number. 
The hysteresis suddenly disappears for Fr less than about $400$. We do not have a clear explanation for this drastic behavior yet. However we observe that the height $\zeta$ starts to decrease from $H$ roughly at ${\rm Fr}\lesssim10^3$, therefore, a crossover of viscous and gravitational coiling regime should exist around ${\rm Fr}=400$. From these facts, we expect that the viscous coiling has hysteresis effect for coil-uncoil transition. To verify this, further experiments or direct numerical simulation should be carried out.

The model we proposed in the present Letter could also reproduce recently reported meandering instability of viscous fluid rope falling onto a moving surface \cite{chu} by simply replacing one of the boundary condition in Eq. (\ref{eq:bc_2}) as $q'_x(0)=s$, where $s$ is the constant share rate. 
Further analysis of this problem is our next work and will appear elsewhere in near future.

We would like to thank Dr. Katsuhiko Sato and Dr. Toshihiro Kawakatsu for their critical comments.

%%%%%%%%%%%%%%%% references %%%%%%%%%%%%%%%%%%%

\end{document}